\documentclass[%
 reprint,
 amsmath,amssymb,
 aps,
]{revtex4-2}

\usepackage{subfigure}
\usepackage{tikz}
\usetikzlibrary{arrows.meta}
\usetikzlibrary{decorations.markings}

\usepackage{graphicx}% Include figure files
\usepackage{dcolumn}% Align table columns on decimal point
\usepackage{bm}% bold math
\begin{document}

\preprint{APS/123-QED}

\title{Dirac materials in parallel non-uniform electromagnetic fields generated by SUSY: \\A new class of chiral Planar Hall Effect?}% Force line breaks with \\

\author{Julio Cesar Pérez-Pedraza}
 \email{julio.perez@umich.mx}
\author{Juan D. García-Mu\~noz}%
 \email{juan.domingo.garcia@umich.mx}
\affiliation{Instituto de F{\'i}sica y Matem{\'a}ticas, Universidad Michoacana de San Nicolás de Hidalgo, Edificio C-3, Ciudad Universitaria, Francisco J. M{\'u}jica S/N Col. Fel{\'i}citas del R{\'i}o, 58040 Morelia, Michoac{\'a}n, M{\'e}xico.
}%


\author{A. Raya}
 \email{alfredo.raya@umich.mx}
\affiliation{
 Instituto de F{\'i}sica y Matem{\'a}ticas, Universidad Michoacana de San Nicolás de Hidalgo, Edificio C-3, Ciudad Universitaria, Francisco J. M{\'u}jica S/N Col. Fel{\'i}citas del R{\'i}o, 58040 Morelia, Michoac{\'a}n, M{\'e}xico
 % with \\
}%
\affiliation{
 Centro de Ciencias Exactas - Universidad del Bio-Bio. Avda. Andr\'es Bello 720, Casilla 447, Chillán, Chile.
}%
%\affiliation{%


\date{\today}% It is always \today, today,

\begin{abstract}
Within a Supersymmetric Quantum Mechanics (SUSY-QM) framework, the (3+1) Dirac equation  describing a Dirac material in the presence of external parallel electric and magnetic fields is solved. Considering static but non-uniform electric and magnetic profiles with translational symmetry along the $y$-direction, the Dirac equation is transformed into two decoupled pairs of Schr\"odinger equations, one for each chirality of the fermion fields. Taking trigonometric and hyperbolic profiles for the vector and scalar potentials, respectively, we arrive at SUSY partner P{\"o}schl-Teller-like quantum potentials. Restricting to the conditions of the potentials that support an analytic zero-mode solution, we obtain a nontrivial current density in the same plane where the electric and magnetic fields lie, but perpendicular to both of them, indicating the possibility of realizing the Planar Hall Effect. Furthermore, this non-vanishing current density is the sum of current densities for the left- and right-chiralities, suggesting that the net current is a consequence of chiral symmetry.
% \begin{description}
% Secondary publications and information retrieval purposes.
% You may use the \texttt{description} environment to structure your abstract;
% use the optional argument of the \verb+\item+ command to give the category of each item. 
% \end{description}
\end{abstract}

\keywords{Dirac Equation, Electromagnetic Fields, Valley Planar Hall Effect, Supersymmetric Quantum Mechanics.}%Use showkeys class option if keyword
\maketitle

%\tableofcontents

\section{Introduction}

Understanding the dynamics of electrons under the influence of external electromagnetic fields lies at the very heart of quantum mechanics~\cite{Cohen-Tannoudji:101367,Sakurai:1167961,griffiths:quantum}. Not surprisingly, the influence of external fields results in a plethora of quantum phenomena. Addressing the problem under the most general assumptions regarding field configurations is indeed a hard nut to crack. A number of simplifications have been considered since the early days of the establishment of quantum physics. For example, for static fields, the quantization of electron orbitals in the plane perpendicular to the field lines of a uniform magnetic field, the Landau levels~\cite{Landau1930}, are the cornerstone in the description of several quantum phenomena~\cite{MIRANSKY20151}. Adding an  electric field parallel to the Landau levels plane generates an electric current perpendicular to both fields. For finite size samples, the quantization of the electric resistance is observed in the Quantum Hall effect (QHE)~\cite{doi:10.1143/JPSJ.39.279,Klitzing} whose smoking gun is the development of plateaus in the transverse conductivity of a bi-dimensional sample of a two-dimensional electron gas at integer numbers of the so-called filling factor, which counts the number of Landau levels filled in the sample.  The quantization rule for the conductivity has also been found for non-integer values of the filling factor~\cite{Tsui} as a result of electron-electron interactions which become relevant~\cite{Laughlin}. Even more, in the anomalous quantum Hall effect~\cite{Novoselov2005,Zhang2005,Gusynin2005} which emerges for relativistic-like excitations in graphene and related materials, it is precisely the relativistic nature of the excitations which describes a sort of integer QHE, with a shifted plateaus. This particular example establishes a natural connection of condensed matter and high-energy physics~\cite{Geim2007}. The importance of the QHE can be further established by the close relation of this effect and the superconducting properties of spin liquid systems~\cite{laughlin1988}. Thus, it is natural to expect the response of charged particles to external fields in variants of this effect.

For time varying fields, like those generated during relativistic heavy-ion collisions in RHIC and LHC~\cite{Brandenburg2021}, it was conjectured that the non-trivial nature of the quantum chromodynamics vacuum could be probed through the generation of a non-dissipative electric current generated by the chiral imbalance of quarks interacting with topological sectors of the gauge field~\cite{Kharzeev:2007jp, KHARZEEV2014133,Fukushima}. Chiral anomaly is responsible for the imbalance in this effect. Nevertheless, a number of thorough measurements in isobaric collisions showed no evidence of the chiral magnetic effect (CME)~\cite{STAR:2021mii, Koch:2016pzl}. An (Abelian) analog of the CME was found in ZrTe$_5$~\cite{Li2016, li2015giant, PhysRevB.93.115414, PhysRevX.5.031023}. In this case, rather than topological configurations of the gauge fields, it is a setup where electric and magnetic fields are parallel which is responsible for the effect.
Remarkably, the chiral anomaly generated by the effect of parallel electric and magnetic fields has also been related to the emergence of a positive longitudinal magnetoconductance~\cite{ZDWLDW2020} in Dirac materials. Along with non-trivial effects of Berry curvature, the chiral anomaly also gives rise to the so-called Planar Hall Effect (PHE)~\cite{NSTT2017} that contrary to the standard QHE setup, in this case the applied current, magnetic field, and the transverse voltage lie coplanarly.

In this work we explore the possibility of realizing the PHE by static but non-uniform electric and magnetic fields, as can be naturally addressed in the context of  $\theta$-electrodynamics (see, for instance, Ref.~\cite{urrutia}). For this purpose, we factorize the four-component Dirac equation under the influence of parallel electric and magnetic fields along the third spatial dimension in terms of two decoupled two-component Weyl-type equation corresponding to each chirality. Each bi-spinor equation can be further factorized in the standard procedure of supersymmetric quantum mechanics. Considering Pöschl-Teller-like  quantum potentials, we look for the conditions of these potentials to support a zero-mode analytic solution. Interestingly, this mode leads to a nontrivial electric current in the same plane of the electric and magnetic field, but perpendicular to both. To present the details of our findings, we have organized the remaining of this article as follows: In Sec.~\ref{Sec:DE} we recall the covariant form of the (3+1) Dirac equation in the presence of external electromagnetic fields. %where we tackle the case of parallel electric and magnetic fields by means of the {\color{blue}SUSY-QM} (briefly described in there). 
Later, in Sec.~\ref{Sec:PT}, we consider a confining example of scalar and vector potentials which leads to a novel chiral Planar Hall Effect. Finally, in Sec.~\ref{Sec:Conclusions} we discuss some aspects and consequences of the aforementioned effect and draw some conclusions.

\section{Dirac equation in electromagnetic fields}\label{Sec:DE}
The Dirac equation describing the motion of free electrons is given by
\begin{equation}\label{E1}
  (i\gamma^{\mu} \partial_{\mu} -m )\Psi_{D}(\mathbf{x}) =0,
\end{equation}
where $\Psi_{D}(\mathbf{x})$ is a four-component spinor, and the Dirac $\gamma^{\mu}$-matrices fulfill the \textit{Clifford} algebra $\lbrace \gamma^{\mu}, \gamma^{\nu} \rbrace =2 g^{\mu\nu}$, with $g^{\mu\nu}=\rm{diag}(1,-1,-1,-1)$ being the (3+1)-dimensional Minkowsky space-time metric. We choose to work in the Dirac representation of these matrices, 
\begin{eqnarray}\label{E2}
    \gamma^0= \beta = 
    \begin{pmatrix}
    1 & 0 \\
    0 & -1 \\
    \end{pmatrix},~ 
    \gamma^i= \beta \alpha_i = 
    \begin{pmatrix}
    0 & \sigma_i \\
    -\sigma_i & 0 \\
    \end{pmatrix},\nonumber \\ \nonumber \\
    \alpha_i=\begin{pmatrix}  0 & \sigma_i \\ \sigma_i & 0 \\   \end{pmatrix},~ 
    \gamma^5= i \gamma^0 \gamma^1 \gamma^2 \gamma^3 = 
    \begin{pmatrix}
    0 & 1 \\
    1 & 0 \\
    \end{pmatrix}.
\end{eqnarray}
It is worth noticing that we are working in natural units $\hbar = c = e = 1$, where $c$ is the speed of light and $e$ is the electron charge. However, in Dirac materials, the Dirac-like equation describing them has a factor directly proportional to the Fermi velocity $v_F$ instead of $c$ (see for example \cite{McCann2013}), thus, the system addressed here is not relativistic. In order to formulate the Dirac equation in the presence of electromagnetic fields, we start by taking the minimal coupling rule $\pi_{\mu} = i \partial_{\mu} - A_{\mu}$ which incorporates the electromagnetic potentials $A_{\mu}=(\Phi,\mathbf{A})$ in the Dirac equation. Thus, Eq. \eqref{E1} becomes
\begin{equation}\label{E3}
   (\gamma^{\mu} \pi_{\mu} -m )\Psi_{D}(\mathbf{x}) =0.
\end{equation}
Considering the massless case, it is easy to show that the squared operator $(\gamma\cdot\pi)^{2}$ can be written as follows 
\begin{equation}\label{E4}
     (\boldsymbol{\gamma} \cdot \boldsymbol{\pi}) ^2 %= \gamma^{\mu}\gamma^{\nu}\pi_{\mu}\pi_{\nu} 
     = \pi^2  + \frac{\sigma^{\mu\nu}}{2} F_{\mu\nu},
\end{equation}
where $\sigma^{\mu\nu} = \frac{i}{2}[\gamma^{\mu}, \gamma^{\nu}]$, and 
\begin{equation}\label{E5}
    F_{\mu\nu} = [\pi_{\mu}, \pi_{\nu}] 
    %\equiv \partial_{\mu} A_{\nu} - \partial_{\nu} A_{\mu}
    =    \begin{pmatrix}
    0 && E_x && E_y && E_z \\
    -E_x && 0 && -B_z && B_y \\
    -E_y && B_z && 0 && -B_x \\
    -E_z && -B_y && B_x && 0
    \end{pmatrix},
\end{equation}
is the electromagnetic field tensor \cite{Margia2010}. Let us analyse each term in Eq. \eqref{E4} separately. First,
\begin{eqnarray} \label{E6}
   \pi^2 &=& %{\color{red}\boldsymbol{\pi} \cdot \boldsymbol{\pi}} = g^{\mu\nu}(i \partial_{\mu}- A_{\mu})(i \partial_{\nu}- A_{\nu}) \nonumber \\
    -\partial_t^2 - i \frac{\partial A_0}{\partial t} -2i A_0 \partial_t + A_0^2 + \partial_j^2 \nonumber \\
   && +i \frac{\partial A_j}{\partial x^j} +2i A_j \partial_j - A_j^2.
\end{eqnarray}
On the other hand, noticing that
\begin{align}
    \frac{\sigma^{0\nu}}{2} F_{0\nu} &= \frac{i}{2} \alpha_i F_{0i}\nonumber\\
    & = \frac{i}{2}(\alpha_1 F_{01}+\alpha_2 F_{02}+\alpha_3 F_{03}), \label{E7}\\
    \frac{\sigma^{ij}}{2} F_{ij} &= -\frac12 \epsilon_{ijk} F_{ij}I\otimes \sigma_k  \nonumber\\
    &= - I \otimes(\sigma_3 F_{12}- \sigma_2 F_{13}+\sigma_1 F_{23}), \label{E8}
\end{align}
where $I = I_{2\times 2}$ is the identity matrix, the second term can be written as
\begin{eqnarray}\label{E9}
    \frac{\sigma^{\mu\nu}}{2} F_{\mu\nu} &=& \frac{i}{2}(\alpha_1 F_{01}+\alpha_2 F_{02}+\alpha_3 F_{03}) \nonumber\\ 
    &-& I\otimes (\sigma_3 F_{12}- \sigma_2 F_{13}+\sigma_1 F_{23}),
    %&\begin{pmatrix}
    %-\sigma_3 F_{12} +\sigma_2 F_{13} -\sigma_1 F_{23}&& i(\sigma_1 F_{01} %+\sigma_2 F_{02} +\sigma_3 F_{03})\\ \\
    %i(\sigma_1 F_{01} +\sigma_2 F_{02} +\sigma_3 F_{03})&&-\sigma_3 F_{12} %+\sigma_2 F_{13} -\sigma_1 F_{23} 
    %\end{pmatrix},
\end{eqnarray}
where we have used the relations $[\gamma^0, \gamma^i] = 2 \alpha_i$, $[\gamma^i, \gamma^j] = i 2 \epsilon_{ijk} \sigma_k \otimes I_{2\times 2}$, $F_{\mu\nu} = -F_{\nu\mu}$ and $\sigma^{\mu\nu} = -\sigma^{\nu\mu} $. Equation \eqref{E4} together with the expressions in Eqs. \eqref{E6} and \eqref{E9} represent all the possible configurations of electromagnetic fields coupled to the Dirac equation describing the dynamics of fermions under its influence. In particular, we are interested in configurations allowing the use of the Supersymetric Quantum Mechanics (SUSY-QM) to solve the eigenvalue problem of the Dirac Hamiltonian. It has been demonstrated that when we only consider a magnetic field perpendicular to the $X-Y$ plane, where fermions are constrained to move, the system can be solved within a SUSY framework (see, for example, Ref.~\cite{Kuru2009}). Analogously, in the case of an electric field pointing along the $X$-direction, SUSY-QM is also a framework useful to find the respective solutions~\cite{Ghosh2017}. On the other hand, from Eq.~\eqref{E9} we observe that a configuration of magnetic and electric fields does not lead to a system of equations which can be solved by a suitable SUSY-QM. Below we study a particular setup  in which the system exhibits supersymmetry.

\subsection{Parallel electromagnetic fields}
Let us consider a static parallel fields configuration, namely, a magnetic field $\mathbf{B} = B(x)\ \mathbf{\hat{z}}$ and an electric field $\mathbf{E} = E(z)\ \mathbf{\hat{z}}$. Working in the Landau gauge, the vector potential generating $\mathbf{B}$ can be chosen as $\mathbf{A} = A(x)\ \mathbf{\hat{y}}$. Moreover, since the rotational of the electric field is equal to zero, this field is generated by  a electric scalar potential $\phi(z)$. Thus, the field strengths turn out to be 
\begin{equation} \label{E10}
    B(x) = \frac{dA(x)}{dx}, \quad E(z) = -\frac{d\phi(z)}{dz}.
\end{equation}
In this physical situation, the operator $\pi^{2}$ is simplified to the following form
\begin{eqnarray} \label{E11}
    \pi^{2} =& -\partial^{2}_{t} - i2\phi(z)\partial_{t} + \phi^{2}(z) + \partial^{2}_{x} + \partial^{2}_{y} + \partial^{2}_{z} \nonumber \\ &+ i2A(x)\partial_{y} - A^{2}(x),
\end{eqnarray}
while the term $\sigma^{\mu\nu}F^{\mu\nu}/2$ is explicitly written as 
\begin{equation} \label{E12}
    \frac{\sigma^{\mu\nu}}{2}F^{\mu\nu} = 
    \begin{pmatrix}
    B(x) && 0 && iE(z) && 0 \\
    0 && -B(x) && 0 && -iE(z) \\
    iE(z) && 0 && B(x) && 0 \\
    0 && -iE(z) && 0 && -B(x)
    \end{pmatrix}.
\end{equation}

On the other hand, we assume the {\it spinor} $\Psi_{D}(t,x,y,z)$, such that $(\gamma\cdot\pi)^{2}\Psi_{D}(t,x,y,z) = 0$, has a standard stationary temporal behavior and since the system exhibits translational symmetry along the $Y$-direction,  we propose $\Psi_{D}(t,x,y,z)$ as follows,
\begin{equation} \label{E13}
    \Psi_{D}(t,x,y,z) = e^{i(\varepsilon t + ky)}\Psi(x,z) = e^{i(\varepsilon t + ky)}
    \begin{pmatrix}
    \psi^{\alpha}_{\uparrow} \\
    \psi^{\beta}_{\downarrow} \\
    \psi^{\beta}_{\uparrow} \\
    \psi^{\alpha}_{\downarrow}
    \end{pmatrix},
\end{equation}
with $\varepsilon$ being the energy of the Dirac electron and $k$ its wavenumber in the $Y$-direction. Then, the time-independent {\it spinor} $\Psi(x,z)$ fulfills an eigenvalue problem that is equivalent to the following coupled system of equations
\begin{widetext}
\begin{equation} \label{E14}
    \left\{\left[\partial_{x}^{2} - (k + A(x))^{2} \pm \frac{dA(x)}{dx}\right] + \left[\partial_{z}^{2} + (\phi(z) + \varepsilon)^{2}\right]\right\}\psi_{\uparrow\downarrow}^{\alpha,\beta} \mp i\frac{d\phi(z)}{dz}\psi_{\uparrow\downarrow}^{\beta,\alpha} = 0,
\end{equation}
\end{widetext}
where the arrows $\uparrow \downarrow$ specify the spin orientation. The above system can be decoupled defining 
\begin{equation} \label{E15}
    \varphi^{\pm}_{\uparrow\downarrow} \equiv \psi^{\alpha}_{\uparrow\downarrow} \pm \psi^{\beta}_{\uparrow\downarrow}, \quad 
\end{equation}
such that we are lead to the  following system of equations
\begin{subequations}
\label{eq:susy_eq}
\begin{eqnarray}
&\left[\left( -\partial_{x}^{2} + (k + A(x))^{2} - \frac{dA(x)}{dx}\right) \right. \nonumber \\ & +  \left.\left(-\partial_{z}^{2} - (\phi(z) + \varepsilon)^{2} \pm i\frac{d\phi(z)}{dz} \right)\right] \varphi^{\pm}_{\uparrow} = 0,\label{susy1}
\end{eqnarray}
\begin{eqnarray}
&\left[\left( -\partial_{x}^{2} + (k + A(x))^{2} + \frac{dA(x)}{dx}\right)  \right. \nonumber \\ & +  \left.\left(-\partial_{z}^{2} - (\phi(z) + \varepsilon)^{2} \mp i\frac{d\phi(z)}{dz}\right)\right] \varphi^{\pm}_{\downarrow} = 0.\label{susy2}
\end{eqnarray}
\end{subequations}
%\begin{align} 
%    &\left\{\left[-\partial_{x}^{2} + (k + A(x))^{2} - \frac{dA(x)}{dx}\right] + \left[-\partial_{z}^{2} - (\phi(z) + \varepsilon)^{2} \pm i\frac{d\phi(z)}{dz}\right]\right\}\varphi^{\pm}_{\uparrow} = 0.\label{E16} \\
%    &\left\{\left[-\partial_{x}^{2} + (k + A(x))^{2} + \frac{dA(x)}{dx}\right] + \left[-\partial_{z}^{2} - (\phi(z) + \varepsilon)^{2} \mp i\frac{d\phi(z)}{dz}\right]\right\}\varphi^{\pm}_{\downarrow} = 0. \label{E17}
%\end{align}
%%%%%%%%%%%%%%%%%% EM fields figure %%%%%%%%%%%%%%%%%%%%%%%%%%%%%5%%%%%
\begin{figure*}[ht]
    \centering
    \begin{subfigure}{}
        \centering
         \includegraphics[scale=1]{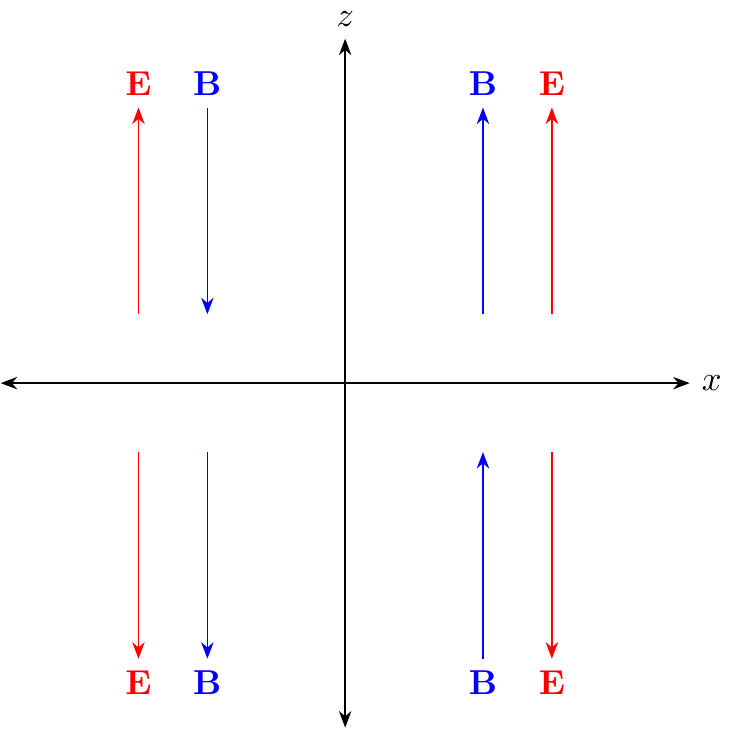}
    \end{subfigure}
    \begin{subfigure}{}
        \centering
        \includegraphics[scale=0.4]{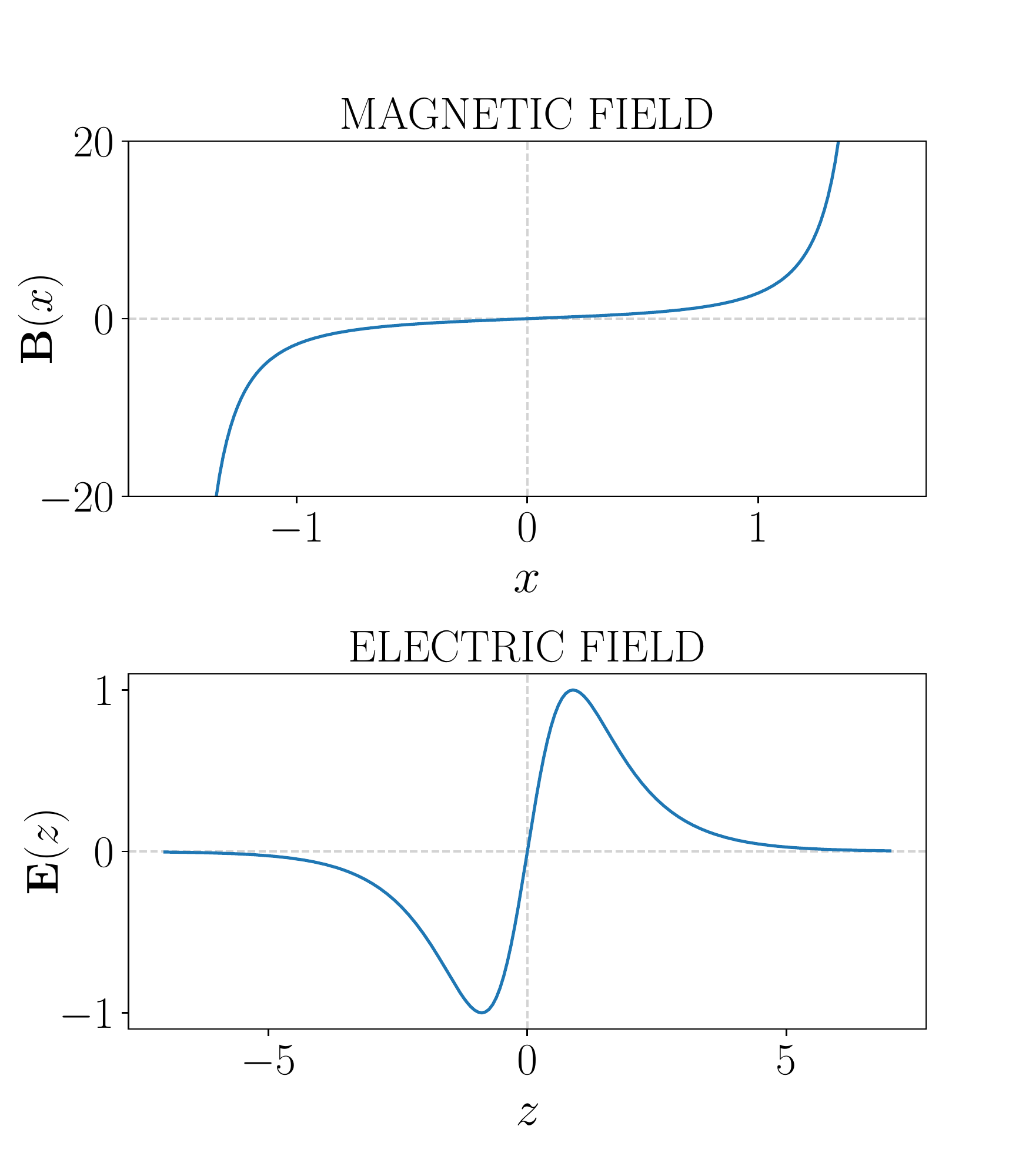}
    \end{subfigure}
 \caption{A sketch of the alignment of the fields in the plane X-Z (left). The electromagnetic fields generated by the potentials in Eq.~\eqref{E25} (right).  The scale of the graphs is set by parameters $E_0= 2.0$, $B_0 = 1.0$, $\mu= 1.0$ and $\nu= 1.0$.}
 \label{F1}
\end{figure*}
Before proceeding, let us briefly review the SUSY-QM framework, in which two Schr{\"o}dinger-like Hamiltonians $H^{\pm}$ are intertwined by means of the operational relation 
\begin{equation} \label{EI1}
    H^{+}L^{-} = L^{-}H^{-},
\end{equation}
with $L^{-}$ being the intertwining operator given by 
\begin{equation} \label{EI2}
    L^{-} = \frac{d}{dx} + w(x),
\end{equation}
where $w(x)$ is a real function referred to as the superpotential. Within, the SUSY-QM framework, the partner potentials $V^{\pm}$ associated to the intertwined Hamiltonians $H^{\pm}$ can be written in terms of the superpotential function as follows
\begin{equation}\label{EI3}
    V^{\pm} = w^{2}(x) \pm w'(x).
\end{equation}
Consequently, both Hamiltonians have an isospectral part and their eigenfunctions are linked through the supersymmetric transformation defined by the intertwining operator~\cite{Gangopadhyaya2018,Junker2019,Fernandez2019}. The SUSY algorithm has been successfully applied to solve the Dirac equation \cite{Ghosh_2009,Ghosh_2021}, in particular, when this equation describes Dirac materials such as graphene~\cite{Kuru2009,FGO2020,Juan2021,FG2022}. 

Defining the Schr{\"o}dinger-like Hamiltonians 
\begin{subequations}
\label{eq:Hamiltonians}
\begin{eqnarray}
    H^{\pm}_{A} = -\partial_{x}^{2} + (k + A(x))^{2} \pm \frac{dA(x)}{dx}, \label{H1}
\end{eqnarray}
\begin{eqnarray}
    H^{\pm}_{\phi} = -\partial_{z}^{2} + [i(\phi(z) + \varepsilon)]^{2} \pm i\frac{d\phi(z)}{dz}, \label{H2}
\end{eqnarray}
\end{subequations}
which are first-order supersymmetric partners, respectively, the corresponding SUSY transformations can be defined by  the superpotentials
\begin{equation} \label{E19}
    \text{w}_{A} = k + A(x),\quad \text{w}_{\phi} = i(\varepsilon + \phi(z)).
\end{equation}
Thus, with the aid of the SUSY algorithm, the system of equations~(\ref{eq:susy_eq}) can be written as
\begin{subequations}\label{E22}
\begin{equation} 
    \left[H^{-}_{A} + H^{\pm}_{\phi}\right]\varphi^{\pm}_{\uparrow} = 0,
\end{equation}
\begin{equation}
    \left[H^{+}_{A} + H^{\mp}_{\phi}\right]\varphi^{\pm}_{\downarrow} = 0.
\end{equation}
\end{subequations}
Expressions above imply that we can construct solutions of the form
\begin{equation}\label{EE1}
    \varphi^{\pm}_{\uparrow} = \chi^{-}_{\uparrow}(x)\zeta^{\pm}_{\uparrow}(z),\quad \varphi^{\pm}_{\downarrow} = \chi^{+}_{\downarrow}(x)\zeta^{\mp}_{\downarrow}(z),
\end{equation}
satisfying each one an eigenvalue equation of the form
\begin{equation} \label{E23}
H^{\pm}_{A}\chi^{\pm}_{\uparrow\downarrow} = \varepsilon_{A}\chi^{\pm}_{\uparrow\downarrow}, \quad H^{\pm}_{\phi}\zeta^{\pm}_{\uparrow\downarrow} = \varepsilon_{\phi}\zeta^{\pm}_{\uparrow\downarrow}.
\end{equation}
Hence, we obtain a relation between the energies $\varepsilon_{A}$ and $\varepsilon_{\phi}$, given by 
\begin{equation} \label{E24}
    \varepsilon_{A} = - \varepsilon_{\phi}.
\end{equation}
By solving the time-independent Schr{\"o}dinger-like equations~\eqref{E23} such that the constraint in Eq.~\eqref{E24} is fulfilled, it is possible to determine the {\it spinor} $\Psi_{D}$ satisfying the massless Dirac equation. It is worth mentioning the potentials $V_{\phi}^{\pm}(z)$ corresponding to the Hamiltonians in Eq.~\eqref{H2} are complex. Since we are looking for real energy eigenvalues, care must be paid in the choice of the electromagnetic profiles. In the following section we discuss an example of electromagnetic fields which lead to solvable Schr{\"o}dinger-like potentials. We explicitly explore their analytic solutions.  

\section{Confining case: P{\"o}schl-Teller-like potentials}\label{Sec:PT}

In order to determine the bound states of the Hamiltonians in Eq.~\eqref{E23}, let us take electromagnetic potentials of the form
\begin{equation}\label{E25}
    A(x)= \frac{B_{0}}{\nu}\sec(\nu x),\ -\frac{\pi}{2}<\nu x< \frac{\pi}{2};~ \phi(z) = \frac{E_{0}}{\mu}\text{sech}(\mu z).
\end{equation}
In Fig.~\ref{F1} we show the corresponding electric and magnetic fields produced by these potentials. 

\begin{figure*}[ht]
 \includegraphics[width=0.8\linewidth]{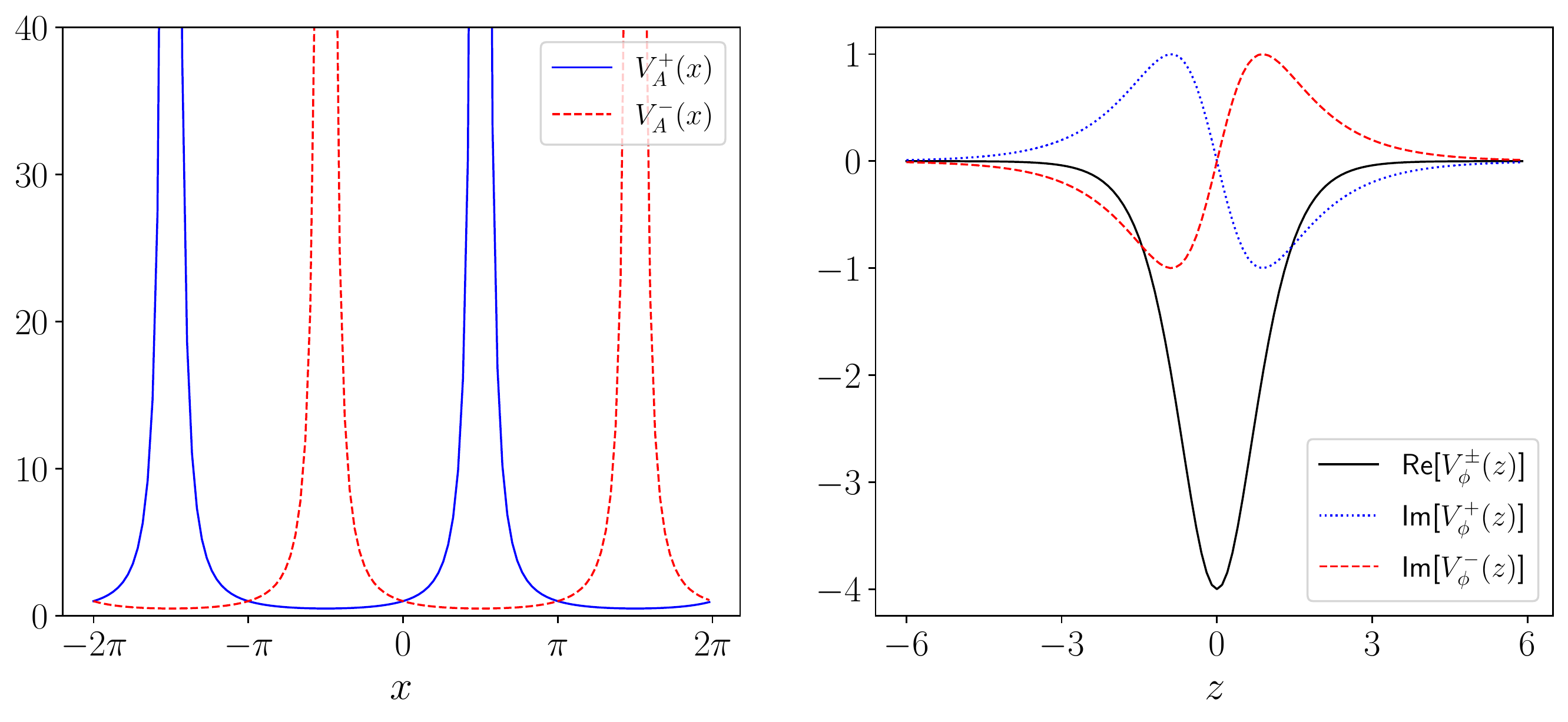}
 \caption{Plot of the P\"osch-Teller-like SUSY partner potentials $V^{\pm}_{A}(x)$ (left). Real and imaginary parts of the potentials $V^{\pm}_{\phi}(z)$ (right).  The scale of the graphs is set by parameters  $E_0= 2.0$, $B_0 = 1.0$, $\mu= 1.0$, $\nu= 1.0$ and $\varepsilon = k = 0$.}
 \label{F2}
\end{figure*}

With the definitions above, from Eqs.~\eqref{EI3} and \eqref{E19}, the SUSY partner potentials are
\begin{subequations}
\label{E26}
\begin{eqnarray}
     V_{A}^{\pm}(x) = &k^2 + 2kD_{A}\sec(\nu x) + D_{A}^{2}\sec^{2}(\nu x) \nonumber \\ &\pm \nu D_{A}\sec(\nu x)\tan(\nu x), 
\end{eqnarray}
\begin{eqnarray}
     V_{\phi}^{\pm}(z) = &-\varepsilon^{2} - 2\varepsilon D_{\phi}\text{sech}(\mu z) - D_{\phi}^{2}\text{sech}^{2}(\mu z) \nonumber \\ &\mp i\mu D_{\phi}\text{sech}(\mu z)\tanh(\mu z),  \label{V2}
\end{eqnarray}   
\end{subequations}
where $D_{A}=B_{0}/\nu$, $D_{\phi}=E_{0}/\mu$.  Notice that the complex potentials in Eq.~\eqref{V2}, taking $\varepsilon = 0$, are particular cases of the so-called pseudo-Hermitian operators with real energy eigenvalues \cite{Mostafazadeh2002,Mostafazadeh2002b}. It can be seen that both potentials have a similar form. Thus, we focus on getting the solutions of $V^{\pm}_{\phi}$; the potentials $V^{\pm}_{A}$ can be solved analogously. In Fig.~\ref{F2}, we plot the SUSY partner potentials in Eq.~\eqref{E26}.
%%%%%%%%%%%%%%%%%% Potentials figure %%%%%%%%%%%%%%%%%%%%%%%%%%%%%5%%%%%
In order to solve the eigenvalue equation of the Hamiltonians $H^{\pm}_{\phi}$, we perform the change of variable $u=i\sinh(\mu z)$, from which we obtain that
\begin{widetext}
\begin{equation}\label{E27} 
  \left[\mu^{2}(1-u^{2})\frac{d^{2}}{du^{2}} -\mu^{2}u\frac{d}{du} - \varepsilon^{2} - 2\frac{\varepsilon D_{\phi}}{\sqrt{1-u^{2}}} - \frac{D_{\phi}^{2}}{1-u^{2}} \mp \mu D_{\phi}\frac{u}{1-u^{2}} - \varepsilon_{\phi}\right]\zeta^{\pm}(u)=0.   
\end{equation}
\end{widetext}
Considering $\varepsilon = 0$, these differential equations lead to the Jacobi equation. We then propose the wave functions of the form $\zeta^{\pm}(u)=(1-u)^{a}(1+u)^{b}f^{\pm}(u)$. Thus, the differential equations~\eqref{E27} can be written as 
\begin{widetext}
\begin{equation}\label{E28}
\begin{aligned}
    &\left\{(1-u^{2})\frac{d^{2}}{du^{2}}+2[b-a-(b+a+1/2)u]\frac{d}{du}-b(b-1)-a(a-1)-b-a-2ab-\frac{\varepsilon_{\phi}}{\mu^{2}}\right.\\&+\left.\left[2a(a-1)-2b(b-1)-b+a\mp \frac{D_{\phi}}{\mu}\right]\frac{u}{1-u^{2}}+\left[2b(b-1)+2a(a-1)+b+a-\frac{D_{\phi}^{2}}{\mu^{2}}\right]\frac{1}{1-u^{2}}\right\}f^{\pm}(u)=0.
\end{aligned}
\end{equation}
\end{widetext}
Equation~\eqref{E28} accepts as solutions the Jacobi polynomials with complex argument $P_{n}^{(\alpha,\beta)}(u)$  as long as the following conditions are accomplished:
\begin{eqnarray}\label{E29}
    2(b-a)=\beta-\alpha, \nonumber\\
    2(a+b+1/2)= \alpha+\beta+2, \nonumber\\
    -(b+a)^{2}-\frac{\varepsilon_{\phi}}{\mu^{2}}=n(n+\alpha+\beta+1), \\
    (a-b)[2(a+b)-1]=\pm S_{\phi}, \nonumber\\
    2(a^{2}+b^{2})-(a+b)=S^{2}_{\phi}, \nonumber
\end{eqnarray}
with $S_{\phi} = D_{\phi}/\mu$. Once  the above requirements  fulfilled, the eigenfunctions of the Hamiltonians $H^{\pm}_{\phi}$ are given by
\begin{equation}\label{E30}
\begin{aligned}
    \zeta^{\pm}_{n}(u)&=(1-u)^{\frac{1}{4}\left(1-\sqrt{1+4S_{\phi}(S_{\phi}\pm 1)}\right)}\\
    &\times(1+u)^{\frac{1}{4}\left(1-\sqrt{1+4S_{\phi}(S_{\phi}\mp 1)}\right)}\\
    &\times P_{n}^{\left(-\frac{1}{2}\sqrt{1+4S_{\phi}(S_{\phi}\pm 1)},-\frac{1}{2}\sqrt{1+4S_{\phi}(S_{\phi}\mp 1)}\right)}(u).
    \end{aligned}
\end{equation}
Boundary conditions imply that $|S_{\phi}|>1$ and a finite discrete spectrum since the number of square-integrable functions is bounded by $Q_{\phi}-1\geq n$, where $$Q_{\phi}=\frac{\sqrt{1+4S_{\phi}(S_{\phi}+1)}+\sqrt{1+4S_{\phi}(S_{\phi}-1)}}{2}.$$ It is worth-noticing that if $S_{\phi}$ is not an integer, the inequality is strict. Thus, $Q_{\phi}-1$ is a upper value of the number of levels and it is necessary check the square-integrablility of each one of the eigenfunctions. Furthermore, the corresponding eigenvalues are
\begin{equation} \label{E31}
    \varepsilon_{\phi} = -\mu^{2}\left[n(n-Q_{\phi}+1)+\frac{(1-Q_{\phi})^{2}}{4}\right].
\end{equation}

On the other hand, as we mentioned earlier, when $k=0$, the solutions of the Hamiltonians $H_{A}^{\pm}$ can be found by a similar process. Their corresponding eigenfunctions can be written as 
\begin{equation} \label{E32}
    \begin{aligned}
    \chi^{\pm}_{n}(u)&=(1-u)^{\frac{1}{4}\left(1+\sqrt{1+4S_{A}(S_{A}\pm 1)}\right)}\\
    &\times(1+u)^{\frac{1}{4}\left(1+\sqrt{1+4S_{A}(S_{A}\mp 1)}\right)}\\
    &\times P_{n}^{\left(\frac{1}{2}\sqrt{1+4S_{A}(S_{A}\pm 1)},\frac{1}{2}\sqrt{1+4S_{A}(S_{A}\mp 1)}\right)}(u),
    \end{aligned}
\end{equation}
with $u=\sin(\nu x)$ and $S_{A}=D_{A}/\nu$. Boundary conditions are satisfied if $S_{A}(S_{A}\pm 1)>-1/4$. Moreover, the spectrum is infinite discrete, whilst the energy eigenvalues are given by
\begin{equation} \label{E33}
    \varepsilon_{A}=\nu^{2}\left[n(n+Q_{A}+1)+\frac{(1+Q_{A})^{2}}{4}\right],
\end{equation}
where $$Q_{A} = \frac{\sqrt{1+4S_{A}(S_{A}+1)}+\sqrt{1+4S_{A}(S_{A}-1)}}{2}.$$

The complete system, composed by both Hamiltonians including the magnetic and electric potentials (Eq.\eqref{E22}), fulfills the relation for the energies $\epsilon_{A} = - \epsilon_{\phi}$ only when $n=0$, resulting in the condition $Q_{A}=-1 \pm (\mu/\nu)|Q_{\phi}-1|$, which relates both field strengths ($B_0$, $E_0$) and concentrations ($\mu$, $\nu$). Hence, we have a {\it spinor} $\Psi_{0}(x,z)$ with energy eigenvalue $\varepsilon = 0$ and wavenumber $k=0$. To find solutions involving non-zero energies and wavenumbers, the potentials in Eq.~\eqref{E26} should be solved in general, which is a highly non trivial task. Finally, in terms of the functions $\chi^{\pm}_{0}(x)$ and $\zeta^{\pm}_{0}(z)$, the zero-mode {\it spinor} $\Psi_{0}(x,z)$ turns out to be
\begin{equation} \label{EX1}
    \Psi_{0}(x,z) = \frac{1}{2}
    \begin{pmatrix}
        \chi^{-}_{0}(x)\left[\zeta^{+}_{0}(z) + \zeta^{-}_{0}(z)\right] \\
        \chi^{+}_{0}(x)\left[\zeta^{-}_{0}(z) - \zeta^{+}_{0}(z)\right] \\
        \chi^{-}_{0}(x)\left[\zeta^{+}_{0}(z) - \zeta^{-}_{0}(z)\right]\\
        \chi^{+}_{0}(x)\left[\zeta^{+}_{0}(z) + \zeta^{-}_{0}(z)\right]
    \end{pmatrix}.
\end{equation}
In order to realize the fermion behavior of this zero-mode, we are interested in exploring the corresponding probability and current densities. Furthermore, the chiral-decomposition of the spinor above offers a natural form to understand its dynamics in this field configuration.     

\subsection{Probability and current densities}
To obtain a visualization of the behavior of the particles in the system, we calculate the total probability density
\begin{equation} \label{E34}
    \rho = |\Psi_D(t,x,y,z)|^2 = \Psi^{\dagger}(x,z)\Psi(x,z),
\end{equation}
and the probability current densities given by 
\begin{equation}\label{E35}
    j_i = \Psi_D^{\dagger}(t,x,y,z)\alpha_i \Psi_D(t,x,y,z) = \Psi^{\dagger}(x,z)\alpha_i \Psi(x,z),
\end{equation}
where $\alpha_i=1,2,3$ are defined in Eq.~\eqref{E2}. For the {\it spinor} in Eq.~\eqref{E13} these quantities have the following forms
\begin{equation} \label{EX2}
\begin{aligned}
    &\rho = \frac{1}{2} \left[|\psi^{\alpha}_{\uparrow}|^2 +|\psi^{\alpha}_{\downarrow}|^2 + |\psi^{\beta}_{\uparrow}|^2 + |\psi^{\beta}_{\downarrow}|^2\right],\\
    &j_{x} = \rm{Re} \left[ \bar{\psi}^{\alpha}_{\uparrow}  \psi^{\alpha}_{\downarrow}  + \bar{\psi}^{\beta}_{\uparrow}  \psi^{\beta}_{\downarrow}  \right],\\
    &j_{y} = \rm{Im} \left[ \bar{\psi}^{\alpha}_{\uparrow}  \psi^{\alpha}_{\downarrow}  + \bar{\psi}^{\beta}_{\uparrow}  \psi^{\beta}_{\downarrow}  \right],\\
    &j_{z} = \rm{Re} \left[ \bar{\psi}^{\alpha}_{\uparrow} \psi^{\beta}_{\uparrow} - \bar{\psi}^{\alpha}_{\downarrow} \psi^{\beta}_{\downarrow} \right].
    \end{aligned}
\end{equation}
%%%%%%%%%%%%%%%%%% Rho-Jx figure %%%%%%%%%%%%%%%%%%%%%%%%%%%%%5%%%%%%%%
\begin{figure}[ht]
 \includegraphics[width=\linewidth]{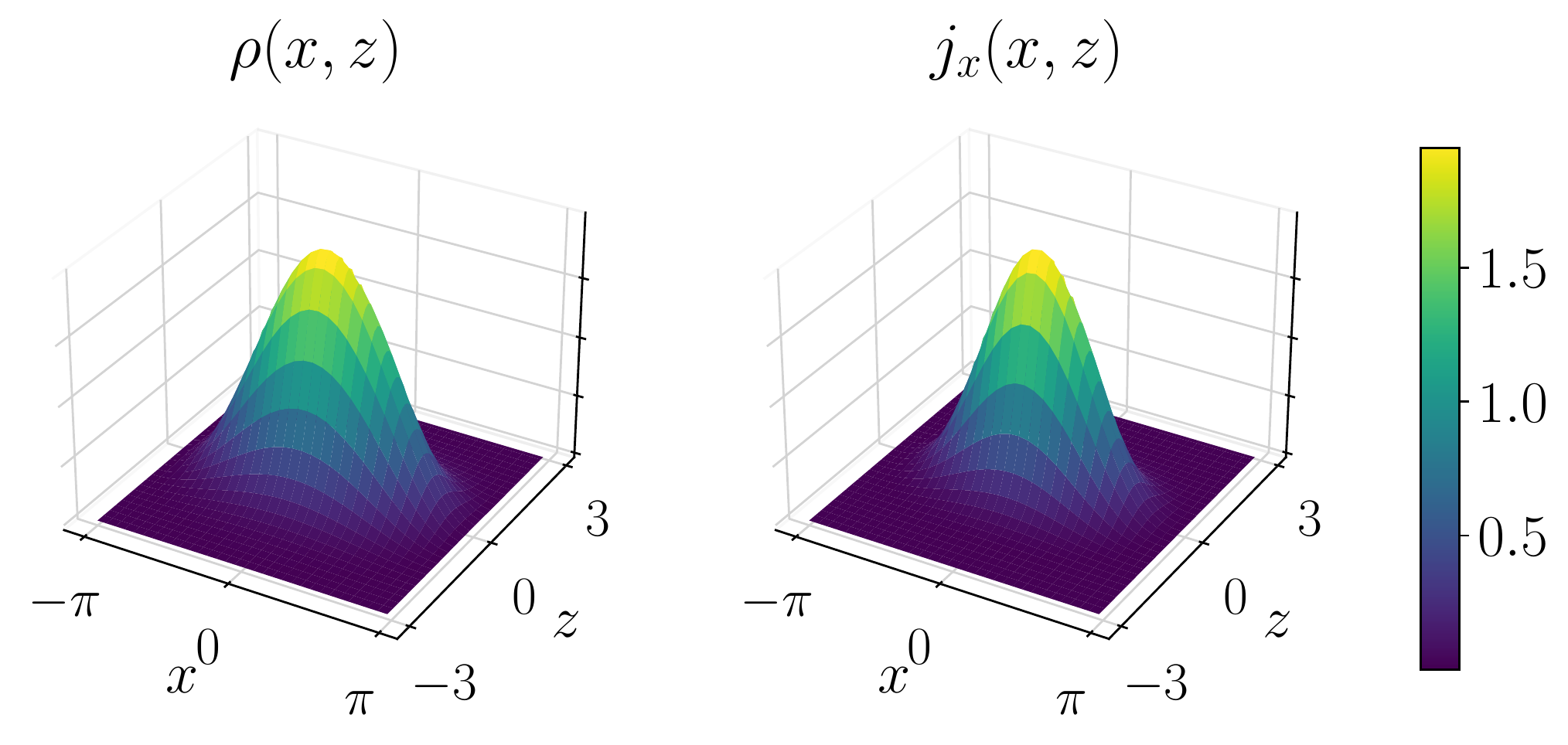}
 \caption{Probability and current densities for the zero-mode written in Eq.~\eqref{EX1}. Note that the parameters $E_0$ and $B_0$ modify the amplitude of the curves, whereas the parameters $\mu$ and $\nu$ expand or concentrate the curves for big and small values, respectively. In these plots the scale is set by the parameters   $E_0= 2.0$, $B_0 = 1.0$, $\mu= 1.0$ and $\nu= 1.0$}
 \label{F3}
\end{figure}

As can be seen from Fig.~\ref{F3}, for the zero-mode {\it spinor} in Eq.~\eqref{EX1}, the probability density is almost entirely localized around the origin, whereas the only non-zero current density points in $X$-direction (we do not show the $Y-$ and $Z-$components of the current density as they exactly vanish).  

\begin{figure}[ht]
 \includegraphics[width=\linewidth]{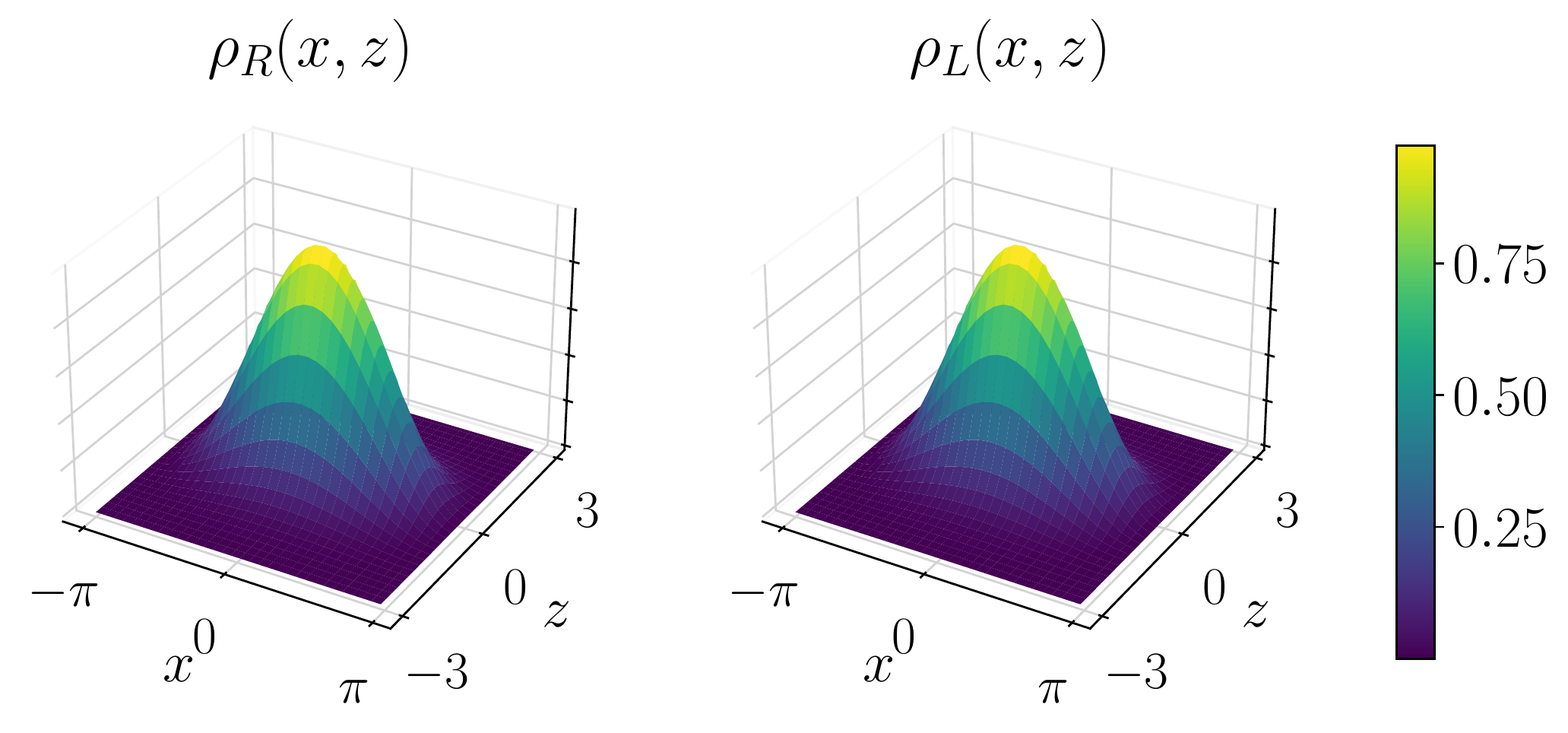}
 \caption{Probability densities for the L- and R-handed {\it spinors} in Eq.~\eqref{E36},  corresponding to the zero-mode {\it spinor} in Eq.~\eqref{EX1}. The scale of the plot is set by the parameters $E_0= 2.0$, $B_0 = 1.0$, $\mu= 1.0$ and $\nu= 1.0$.}
 \label{F4}
\end{figure}

In order to have a better insight of the dynamics, let us look at the chiral decomposition  of the {\it spinor}. For this purpose, we use the chiral projectors $P_{R,L} = \frac{1}{2}(I\pm \gamma_5)$, which allow us to obtain the left- (L) and right-handed (R) {\it spinors}
\begin{eqnarray} \label{E36}
    \Psi_{D;R,L}(t,x,y,z) =& e^{i(\varepsilon t + ky)}\Psi_{R,L}(x,z) \nonumber \\ =& \frac{1}{2} e^{i(\varepsilon t + ky)}
    \begin{pmatrix}
    \psi^{\alpha}_{\uparrow} \pm \psi^{\beta}_{\uparrow} \\
    \psi^{\beta}_{\downarrow} \pm \psi^{\alpha}_{\downarrow} \\
    \psi^{\beta}_{\uparrow} \pm \psi^{\alpha}_{\uparrow} \\
    \psi^{\alpha}_{\downarrow} \pm \psi^{\beta}_{\downarrow} 
    \end{pmatrix},
\end{eqnarray} 
where the down-sign (up-sign) is chosen for the L(R)-handed {\it spinor}. Note that in Eq.~\eqref{E36} we are suppressing the $x$- and $z$-dependence of the {\it spinor} components to simplify the notation. Chiral projections of the probability density (shown in Fig.~\ref{F4}) are
%%%%%%%%%%%%%%%%% Rho-RL figure %%%%%%%%%%%%%%%%%%%%%%%%%%%%%5%%%%%%%%
%%%%%%%%%%%%%%%%%%%%%%%%%%%%%%%%%%%%%%%%%%%%%%%%%%%%%%
%%%% Figura de las tres corrientes quirales
\begin{figure*}[ht]
        \centering
        \begin{subfigure}
            \centering
            %\caption{SMALL PIC}
            \includegraphics[width=0.75\linewidth]{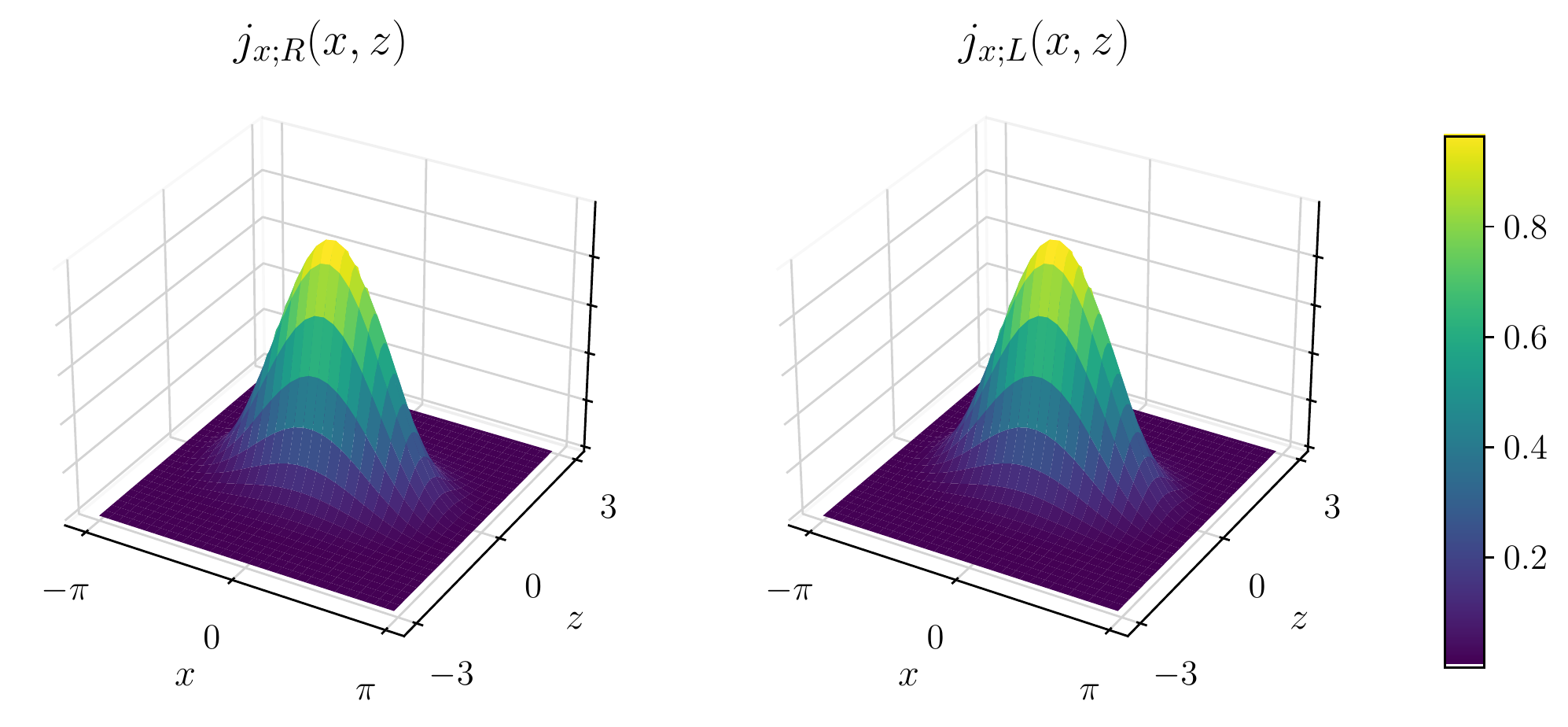}
            \begin{minipage}{.1cm}
            \vfill
            \end{minipage}
        \end{subfigure} 
        \begin{subfigure}
            \centering
            %\caption{SMALL PIC}
            \includegraphics[width=0.75\linewidth]{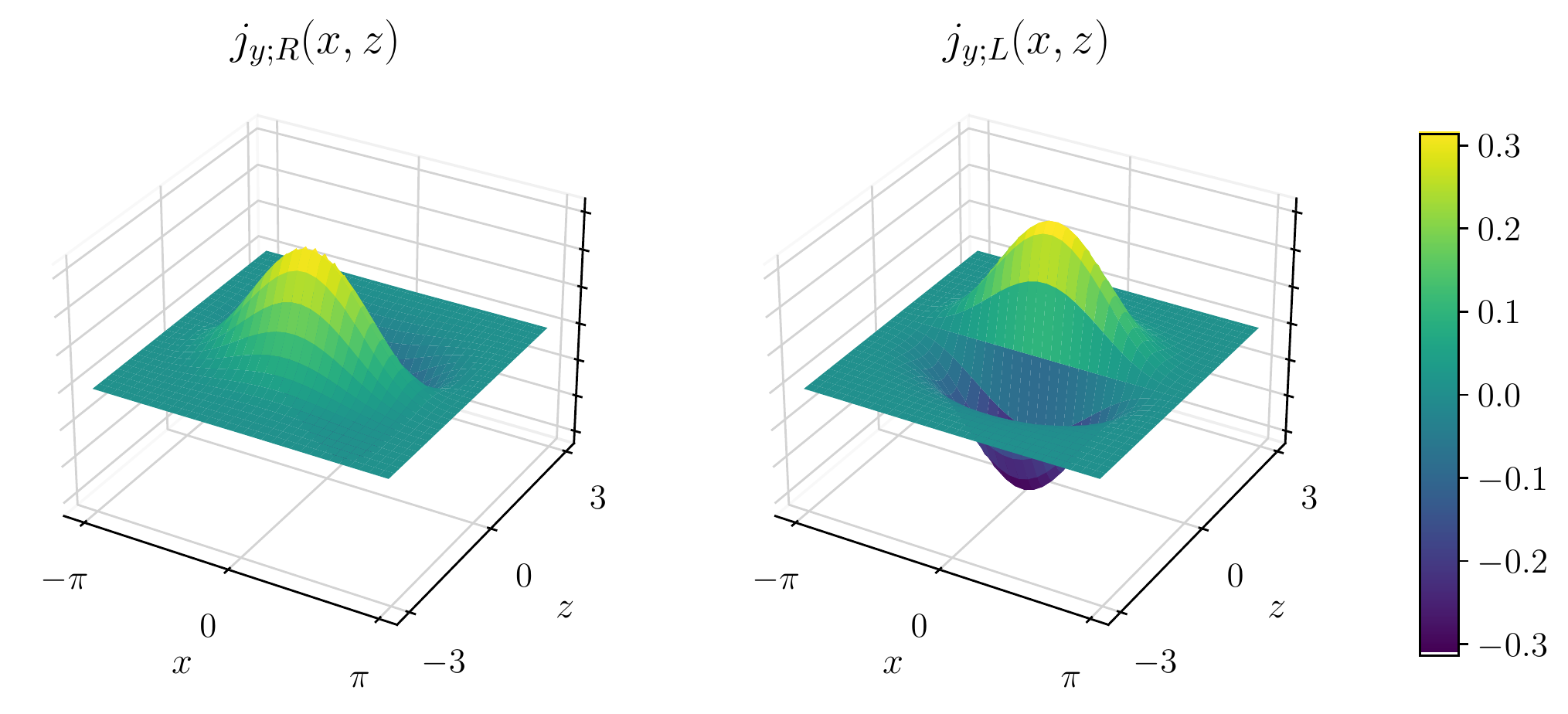}
            \begin{minipage}{.1cm}
            \vfill
            \end{minipage}
        \end{subfigure} 
        \begin{subfigure}
            \centering
            %\caption{SMALL PIC}
            \includegraphics[width=0.75\linewidth]{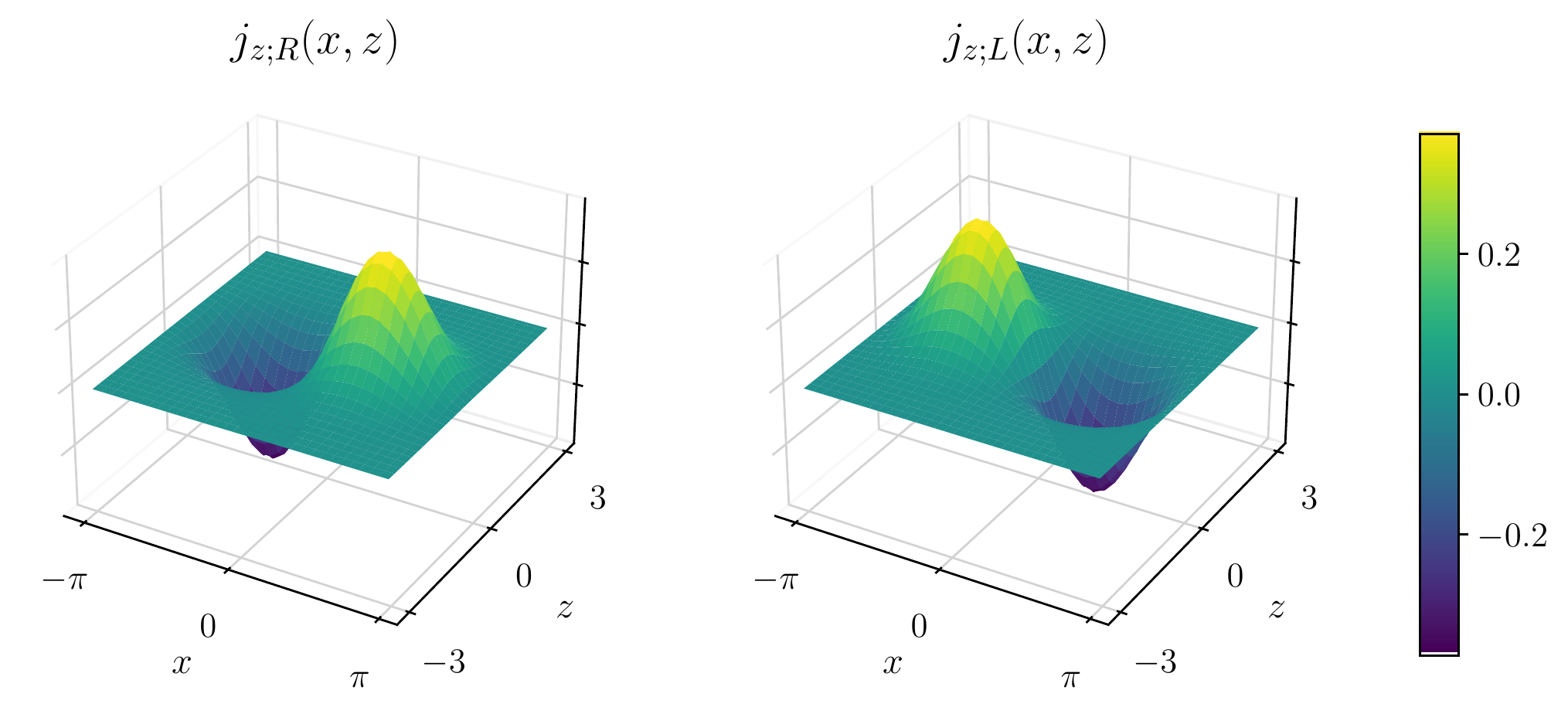}
            \begin{minipage}{.1cm}
            \vfill
            \end{minipage}
        \end{subfigure} 
        \caption{R- and L-chiral probability current densities for the {\it spinors} in Eq.~\eqref{E36} corresponding to the zero-mode of the system. In $y$- and $z$-directions the R- and L-chiral components are exactly equal but with opposite sign, resulting in null-total current densities. While, in $x$-direction both chiral components contribute to the non-zero current density. The parameter values taken are $E_0= 2.0$, $B_0 = 1.0$, $\mu= 1.0$ and $\nu= 1.0$.} \label{F5}
\end{figure*}
\begin{eqnarray}
    \rho_{R,L} = \frac{1}{2} \left[|\psi^{\alpha}_{\uparrow}|^2 +|\psi^{\alpha}_{\downarrow}|^2 + |\psi^{\beta}_{\uparrow}|^2 + |\psi^{\beta}_{\downarrow}|^2\right] \nonumber \\ \pm \rm{Re} \left[ \psi^{*\alpha}_{\phantom{*}\uparrow} \psi^{\beta}_{\uparrow} + \psi^{*\alpha}_{\phantom{*}\downarrow} \psi^{\beta}_{\downarrow} \right],
\end{eqnarray}
and of the current densities (see Fig.~\ref{F5})
\begin{eqnarray}
    j_{x;R,L} &=& \rm{Re} \left[ \psi^{* \alpha}_{\phantom{*}\uparrow} \left( \psi^{\alpha}_{\downarrow} \pm \psi^{\beta}_{\downarrow}\right) + \psi^{*\beta}_{\phantom{*}\uparrow} \left( \psi^{\beta}_{\downarrow} \pm \psi^{\alpha}_{\downarrow}\right)  \right],\\
%\end{equation}
%\begin{equation}
    j_{y;R,L} &=&  \rm{Im} \left[ \psi^{*\alpha}_{\phantom{*}\uparrow} \left( \psi^{\alpha}_{\downarrow} \pm \psi^{\beta}_{\downarrow}\right) + \psi^{*\beta}_{\phantom{*}\uparrow} \left( \psi^{\beta}_{\downarrow} \pm \psi^{\alpha}_{\downarrow}\right)  \right],\\
%\end{equation}
%\begin{eqnarray}
    j_{z;R,L} &=& \rm{Re} \left[ \psi^{*\alpha}_{\phantom{*}\uparrow} \psi^{\beta}_{\uparrow} - \psi^{*\alpha}_{\phantom{*}\downarrow} \psi^{\beta}_{\downarrow} \right]   \nonumber\\
    &\pm& \frac{1}{2} \Big[|\psi^{\alpha}_{\uparrow}|^2+|\psi^{\beta}_{\uparrow}|^2 - |\psi^{\alpha}_{\downarrow}|^2 - |\psi^{\beta}_{\downarrow}|^2\Big] .
\end{eqnarray}
\begin{figure}[ht]
\centering
\includegraphics[scale=1]{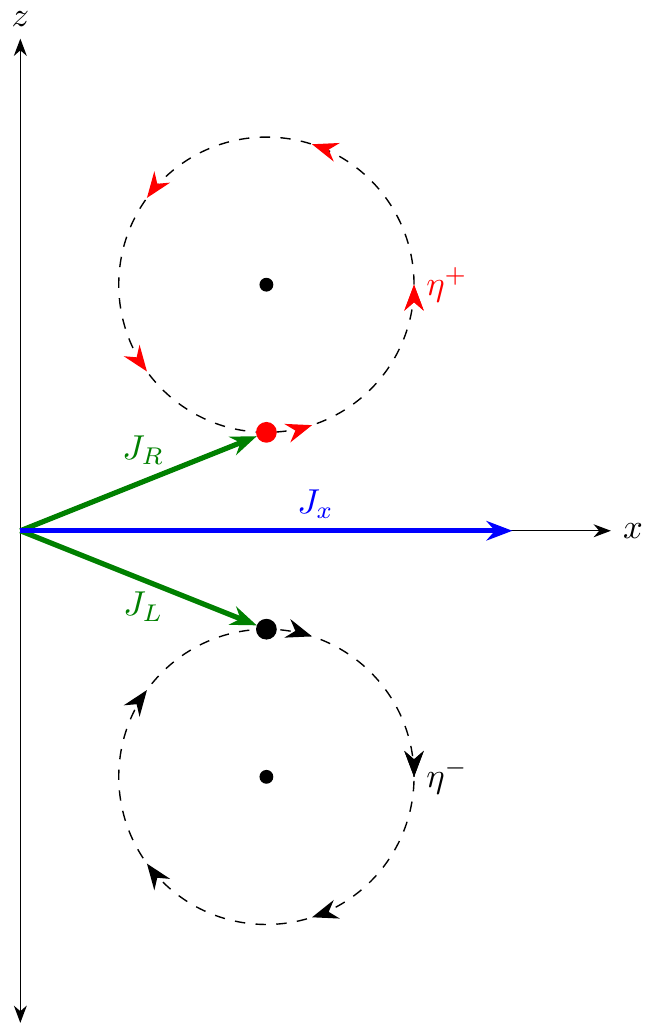}
\caption{A representation of the right and left chiralities. The sum of the corresponding chiral currents is non-zero in $x$-direction.}
\label{F6}
\end{figure}
 A straightforward calculation verifies that the total current densities in Eq.~\eqref{EX2} are the sum of the respective L- and R-handed currents, i.e., $j_i = j_{i;R} + j_{i;L}$. From Fig.~\ref{F5},  we can observe that the L- and R-handed components of the $y-$ and $z-$current densities cancel out (are opposite), whereas in the $X-$direction, they are exactly the same so that the sum  is nonvanishing. From Fig.~\ref{F4}, we note the quantity of particles with L- and R-chiralities is the same (chiral symmetry is conserved). Thus, the continuity equation for the chiral density $\rho_{5} = \rho_{R} - \rho_{L}$ is reduced to
 \begin{equation}\label{chiralEq}
        \nabla\cdot J_{5} = \frac{1}{2\pi^{2}}E(z)B(x),
 \end{equation}
 where $J_{5}$ is the probability current density associated to $\rho_{5}$ \cite{GSDNT2020,LYQ2013}. Taking into account the electromagnetic profiles given in this work, the volume integral of the right-side of Eq.~\eqref{chiralEq} vanishes. Hence, the current density $J_{5}$ describes a circular motion of the chiral particles. It is standard to represent the R-handed chirality ($\eta^{+}$) as a counter-clockwise motion, while L-handed chirality ($\eta^{-}$) is sketched as a clockwise motion. This picture is very useful to gain insight into the behavior of the particles, see Fig.~\ref{F6},  where we can possible observe that the chiral currents sum up in the $X-$direction and cancel out in other directions. In the context of the Dirac materials, given the constrains used to solve the Dirac equation in the previous section, the $X-Z$-plane is the most relevant for the dynamics. Since the non-zero $x$-current density is in the same plane where the electromagnetic fields lie, through our set up we describe a particular case of the Planar Hall Effect (PHE). However, in our case the electromagnetic fields are non-uniform (see Fig.~\ref{F2}), which causes slight differences respect to the standard case where uniform fields are considered~\cite{NSTT2017}. Actually, the inhomogeneous feature of the electromagnetic fields is responsible of the behavior of the particles in the system. Therefore, the current density defining the planar Hall effect is a consequence of the chiral symmetry, which is preserved due to the supersymmetry thereof.  We must mention, in the context of Dirac materials,  chiral symmetry is translated as the valley symmetry, thus, it is also valid to discuss the dynamics in terms of a valley PHE for the system addressed here.
%In the upper quarter-plane both fields point to the same direction ($z-$direction), resulting in an elliptic (in general) movement of the particles (counter-clockwise in this case) referring to a chirality $\eta^+$. On the other hand, In the lower quarter-plane the fields are opposite, thus inverting the direction of the elliptical movement of the particle as well as its chirality ($\eta^-$). 

\section{Conclusions}\label{Sec:Conclusions}
In this article, the (3+1) Dirac equation describing a Dirac material in  the presence of static non-uniform parallel electromagnetic fields is solved within a SUSY-QM framework. In order to determine an exact analytic solution, we address the example of electromagnetic profiles leading to P{\"o}schl-Teller-like quantum potentials. The corresponding zero-mode {\it spinor} is found and its associated probability and current densities obtained. We notice that the current densities vanish in all spatial directions, except for the current along the $X$-direction, which defines a plane in which it lies perpendicularly to the electromagnetic fields. Hence, it is appropriate to assume that a PHE develops in the system dealt here. Furthermore, since the current densities turn out to be written in terms of the left- and right-handed current densities, the effect can be regarded as driven by the chiral symmetry of the system. A chiral-dependent PHE has been shown in inhomogeneous Weyl semi-metals \cite{GSDNT2020}. However, the Dirac material addressed in this article is pristine but under non-uniform external electromagnetic fields. Therefore, this material shows a new class of chiral PHE. We must mention that the electromagnetic profiles in Eq.~\eqref{E25} are tough to realize in the laboratory. Nevertheless, a configuration of pseudo-electromagnetic fields, associated to strains in the material, could become analogous to the system worked here. Such configuration could be feasible in the laboratory through modern strain techniques in Dirac materials, such as scanning tuneling spectroscopy \cite{levy2010strain, klimov2012electromechanical}. Finally, we further remark the localization of the probability density as shown in  Fig.~\ref{F3}. In this sense, it is worth noticing that the considered electromagnetic profiles are setting up two Jackiw-Rebbi (J-R) models for the dynamics along in each direction~\cite{j-r, gorlach2019photonic, jana2019jackiw}. This observation can be seen precisely in Fig.~\ref{F1}. Thus, the wave packet we are considering in this work should be regarded as a two-dimensional soliton of the J-R type. Therefore, our system  could be a promising artificial set up of a wave guide with no energy losses, which could be technologically superb.  

\section*{Acknowledgements}
%JCPP acknowledges support from CONAHCYT through a PhD scholarship. 
We acknowledge enlightening discussions with Ana Julia Mizher and José Alberto Martín Ruiz. Support has been received from CONAHCYT under grant FORDECYT-PRONACES/61533/2020.

\bibliography{dirac_equation1}
\bibliographystyle{unsrt}
%\nocite{NSTT2017,ZDWLDW2020}

\end{document}